\documentclass{INTERSPEECH2023}

\interspeechcameraready 
\usepackage{subcaption}

\title{Personalization for BERT-based Discriminative Speech Recognition Rescoring}
\name{Jari Kolehmainen, Yile Gu, Aditya Gourav, Prashanth Gurunath Shivakumar, Ankur Gandhe, Ariya Rastrow, and Ivan Bulyko}
\address{
  Amazon Alexa
}
\email{jkolehm@amazon.com}

\begin{document}

\maketitle
 
\begin{abstract}

Recognition of personalized content remains a challenge in end-to-end speech recognition. We explore three novel approaches that use personalized content in a neural rescoring step to improve recognition: gazetteers, prompting, and a cross-attention based encoder-decoder model. We use internal de-identified en-US data from interactions with a virtual voice assistant supplemented with personalized named entities to compare these approaches. On a test set with personalized named entities, we show that each of these approaches improves word error rate by over 10\%, against a neural rescoring baseline. We also show that on this test set, natural language prompts can improve word error rate by 7\% without any training and with a marginal loss in generalization. Overall, gazetteers were found to perform the best with a 10\% improvement in word error rate (WER), while also improving WER on a general test set by 1\%.


\end{abstract}


\section{Introduction}
Production automatic speech recognition (ASR) systems often utilizes two passes, where a first pass generates a list of hypotheses, and a second pass rescores the hypotheses to identify likely transcriptions~\cite{xia2017deliberation,hu2020deliberation,hu2021transformer}. For each hypothesis, the second pass models calculate a score (typically, negative log-likelihood) that is combined with the first pass scores to determine the final score~\cite{huang2019empirical,gandhe2020audio}. Scores can be computed using a language model~\cite{huang2019empirical,gandhe2020audio,pandey2022lattention}, a masked language model~\cite{salazar2019masked,liu2021code}, or other hypothesis-level models~\cite{futami2021asr,xu2022rescorebert}. Rescoring performance has been improved by using contextual information in recent literature. This includes audio~\cite{gandhe2020audio,kim2022joint}, lattice information from the first pass~\cite{pandey2022lattention, dai2022latticebart}, or personalized context~\cite{filimonov2020neural,gourav2021personalization}.

Common neural rescoring models~\cite{raju2019scalable,gandhe2020audio,salazar2019masked,futami2021asr,xu2022rescorebert} face challenges with personalized content such as personalized named entities or personal playlists as they cannot adapt to the user's preferences. In particular, this issue manifests itself for homophone names where neither semantic nor acoustic context can distinguish the correct transcription. Personalized and supplemental n-gram models can address these challenges, but they usually lack generalization and overfit to a specific use case, reducing their effectiveness~\cite{gourav2021personalization}. Domain classifiers and other auxiliary models can be used to improve generalization properties. This, however, results in a more complex second pass architecture and higher maintenance costs~\cite{liu2021domain}.  Including textual context directly in a neural model has been shown effective in many tasks such as document classification~\cite{ruggeri2021membert}, language modeling~\cite{zhong2022training}, and acoustic modeling~\cite{sathyendra2022contextual}. In this study, we demonstrate how personalized named entities can be incorporated into a neural rescoring pass.

The literature provides a wide variety of neural modeling approaches that utilize both unstructured and structured contextual information: techniques that use encoded context via a cross-attention framework~\cite{ruggeri2021membert,9687895,sathyendra2022contextual}; retrieval-based approaches~\cite{zhong2022training, borgeaud2022improving}; or approaches that modify the model input via embedding augmentation~\cite{fetahu2022dynamic,ganesan2021n,hu2022gazpne2}. For the last category, it can either be achieved by augmenting token-level input embeddings~\cite{fetahu2022dynamic,hu2022gazpne2} with \textit{gazetteers}.  Alternatively, additional context tokens can be processed as part of the input sequence following certain logic~\cite{ganesan2021n,wen2023hard,shin2020autoprompt,nie2022prompt}. The latter approach is referred to as \textit{prompting}. Prompt embeddings can be trainable~\cite{ganesan2021n,wen2023hard} or discrete optimization can be performed on the existing embeddings to find optimal prompts, without the need to apply conventional model training~\cite{shin2020autoprompt}.

\begin{figure}[!h]
  \centering
     \begin{subfigure}{\linewidth}
         \centering
         \includegraphics[width=\linewidth]{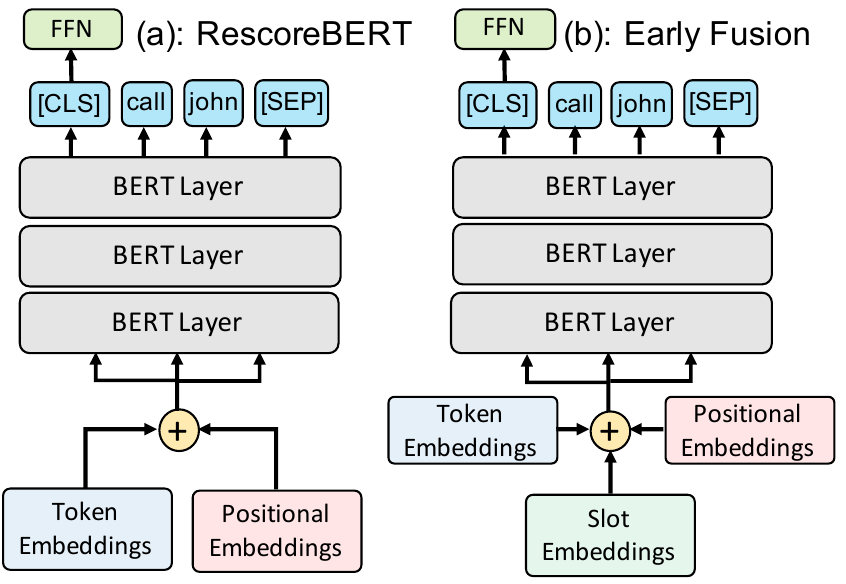}
     \end{subfigure}
     \hfill
     \begin{subfigure}{\linewidth}
         \centering
         \includegraphics[width=0.8\linewidth]{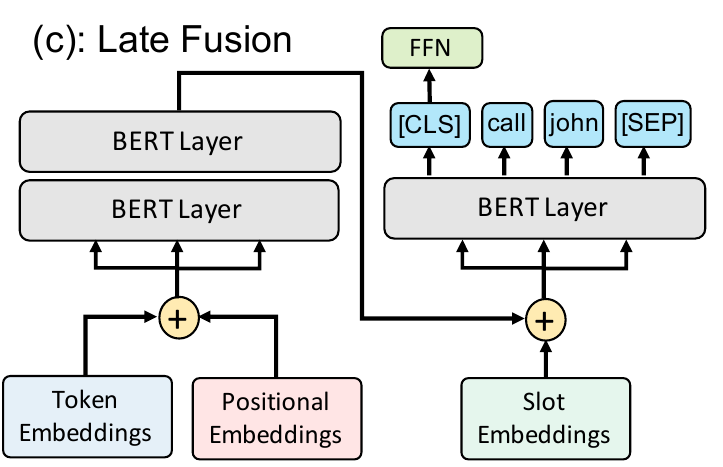}
     \end{subfigure}
    \begin{subfigure}{\linewidth}
         \centering
         \includegraphics[width=\linewidth]{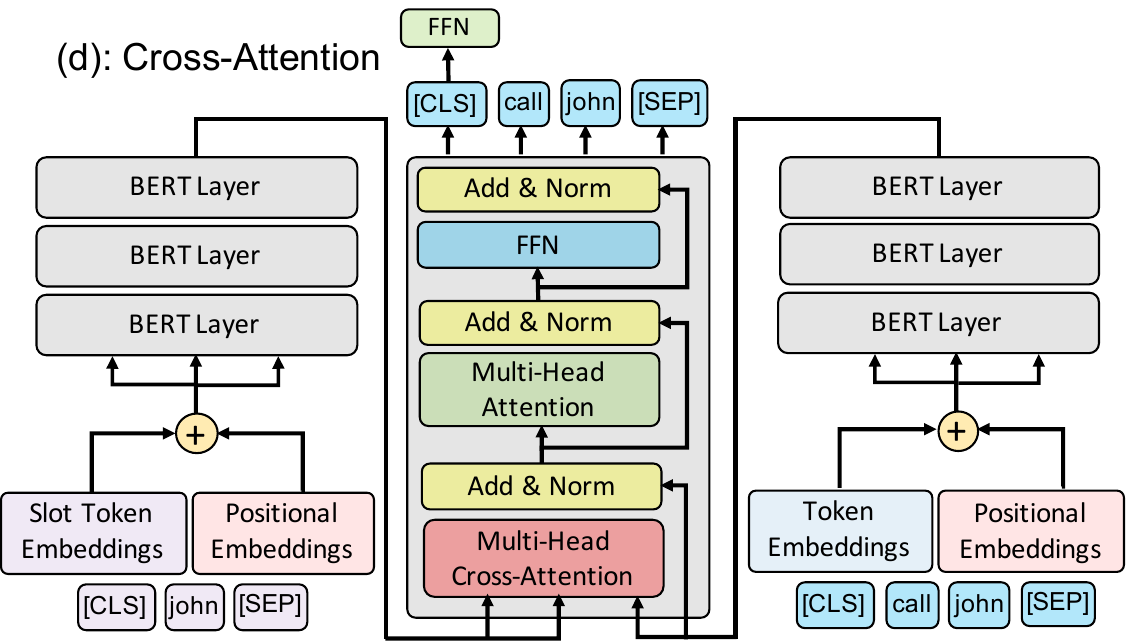}
     \end{subfigure}
    \caption{Illustration of model architectures. (a) RescoreBERT model that attaches a feed-forward network at the classification token embedding of the BERT model. (b) Early fusion gazetteer that adds slot embeddings at the input side of the RescoreBERT model. (c) Late fusion gazetteer that adds the slot embeddings to the inputs of the last RescoreBERT model layer. (d) Cross-attention based encoder-decoder with shared encoders.}
    \label{fig:illustration}
\end{figure}

There are auto-regressive rescoring models~\cite{huang2019empirical,gandhe2020audio,pandey2022lattention} that can be used for rescoring, but bidirectional discriminative models have recently shown superior performance~\cite{futami2021asr, xu2022rescorebert}. In this paper, we use a bidirectional transformer model, called RescoreBERT, as the baseline~\cite{xu2022rescorebert}. RescoreBERT is a rescoring model that predicts a single score for each hypothesis and is trained with a discriminative ASR loss, which directly minimizes the word error rate. The model architecture is summarized in Section \ref{sect:rescoring} and illustrated in Fig. \ref{fig:illustration}a.


To investigate personalization in neural rescoring, we use internal en-US data supplemented with each user's personalized named entities. We focus on three novel rescoring approaches: 1) gazetteers where we add trainable slot embeddings to the vanilla RescoreBERT model; 2) natural language prompting where we include prompts to contextualize the RescoreBERT model; and 3) an encoder-decoder architecture with cross-attention to personalized named entities. We show results for these models in different data scenarios, model sizes, and test sets. This study intends to demonstrate the level of improvement each approach can yield in recognizing personalized content and entities without compromising general recognition performance.


\section{Proposed Approaches}
\label{sect:approaches}

\subsection{Rescoring Model}
\label{sect:rescoring}
We use the recently proposed RescoreBERT as the baseline rescoring model \cite{xu2022rescorebert}. RescoreBERT model uses a pre-trained BERT model with a feed-forward network attached to the classification (CLS) token embedding to predict a single score for each hypothesis (see Fig. \ref{fig:illustration}a). Unlike auto-regressive language models, RescoreBERT scores do not follow a proper probability distribution. In more formal terms, the final score is calculated by,
\begin{equation}
    v_i = \alpha u_i + \beta s_i
    \label{eq:rescore_bert}
\end{equation} where $u_i$ is the first pass score; $\alpha$ and $\beta$ are hyper-parameters (in this study they are $20$ and $1$, respectively); and $s_i$ is the rescoring model score. The rescoring model score is computed by
\begin{equation}
    s_i = f \circ g(E_t(x_{i}) + E_p).
\end{equation}
Where $f$ represents additional feed-forward layers operating on the CLS token embedding.; $g=g_n \circ \cdots \circ g_1$ is a transformer encoder consisting of $n$ layers; $E_t(x_{i})$ and $E_p$ are token and (learnable) positional embeddings, respectively; and $x_{i}$ are hypothesis tokens. In this study, the feed-forward-network $f$ is a linear layer without any activation functions.

The rescoring model is trained using a standard minimum word error rate (MWER) loss calculated on n-best hypotheses from the first-pass: loss~\cite{hori2016minimum,prabhavalkar2018minimum,wang2022improving}:
\begin{equation}
    L = \sum_i (\epsilon_{i} - \overline{\epsilon}) p_{i},
\end{equation} where $\epsilon_{i}$ is the edit distance of $i^{\mathrm{th}}$ hypothesis; $\overline{\epsilon}$ is the mean edit distance of all hypotheses originating from the same utterance; and $p_{i}$ is the normalized posterior distribution of the $i^{\mathrm{th}}$ hypothesis in the n-best, given by the following softmax:
\begin{equation}
    p_{i} = \frac{\exp({-v_{i})}}{\sum_k \exp({-v_{k})}}.
\end{equation}

\subsection{Prompts}
\label{sect:prompts}
We consider a setting where each utterance has a set of tokenized strings $D$ representing personalized named entities. In the prompt approach, RescoreBERT input is concatenated with a natural language prompt if any sub-string (continuous sub-sequence of tokens) in the hypotheses matches an entity in the set $D$. We call this a \textit{match condition}. For example, if [entity] is among the personalized contact names for a particular hypothesis ``call [entity]", then the \textit{match condition} is met. In this study, we did not perform any prompt tuning or prompt engineering and instead used a simple phrase ``as i need to contact [entity]” as the augmented prompt. When there are multiple matched entities within a hypothesis, ``and" is used in the prompt to separate them.


\subsection{Gazetteers}
For gazetteers, we process the input tokens $x_i$ by tagging each token that matches an element in $D$. Only full matches are tagged and partial matches are ignored. The result is a binary sequence $y_i$, in which zeros indicate a non-match and ones indicate a match. Note that $y_i$ has the same length as the input tokens $x_i$. An embedding layer can convert the slot tags $y_i$ into dense representations, with tagged tokens (i.e. $y_i=1$) generating a learnable embedding and non-tagged tokens (i.e. $y_i=0$) generating a fixed non-learnable zero embedding. We consider two gazetteer variants: early fusion and late fusion, which differ in how the slot embeddings are added to the model.

In early fusion, the slot embeddings are added before the first transformer layer into the token and positional embeddings (see Fig. \ref{fig:illustration}b). The resulting score is computed by
\begin{equation}
    s_{i} = f \circ g(E_t(x_{i}) + E_p + E_s(y_{i})),
    \label{eq:early}
\end{equation} where $E_s$ are the slot embeddings. Note that the slot embeddings are added to the existing embeddings instead of replacing them. This will allow model to learn to distinguish salient words (i.e. ``will")  from actual names.

In late fusion, slot embeddings are added to the last transformer layer (see Fig. \ref{fig:illustration}c), the score is computed using:
\begin{equation}
    s_i = f \circ g_n \left(E_s + g_{n-1} \circ \cdots \circ g_1(E_t + E_p)  \right).
    \label{eq:late}
\end{equation}
The slot embeddings $E_S$ are zero when none of the tokens are tagged, and the score is the same as the baseline RescoreBERT in Eq. \eqref{eq:rescore_bert}, for both fusion variants.

\subsection{Cross-Attention Based Encoder-Decoder}
The last personalization approach we considered combines prompting and gazetteers ideas using a cross-attention architecture, as illustrated in Fig. \ref{fig:illustration}d. We use string-matching as before to find all the matching entities from hypotheses that are in $D$. The matched tokenized personalized entities are concatenated with a classification token (CLS) and separated by separation token (SEP) to form slot token sequence $z_i$. Hence, $z_i$ will appear as [CLS] [entity \#1] [SEP] [entity \#2] [SEP] $\cdots$ [SEP]. The slot tokens $z_i$ are then encoded using the same encoder as the hypothesis input tokens $x_i$. Both the hypothesis and the encoded slot embeddings are used as inputs to a non-causal transformer decoder layer to extract a classification token for the scoring feed-forward network. This is more formally expressed by
\begin{eqnarray}
    X_{i} = g(E_t(x_{i}) + E_p) \\
    Z_{i} = g(E_t(z_{i}) + E_p) \\
    s_{i} = f \circ m(Z_{i}, X_{i}),
\end{eqnarray} where $m(\cdot, \cdot)$ is a decoder layer without a causal mask where the first entry represents the keys and values of the multi-head cross-attention and the second entry is for the query (see Fig. \ref{fig:illustration}d). If there are no matching name, the $z_{i}$ is a single classification token. In this study, the decoder layer used the same hyper-parameters as the encoder (e.g. number of attention heads, hidden size, etc.).



\section{Experiments}

\subsection{Data}
We use de-identified far-field en-US datasets gathered from interactions with a virtual voice assistant in this study. Training data is split into two parts: The first part consists of personalized utterances where the ground truth transcription contains a reference to a personalized named entity; a second part consists of randomly selected utterances from usage traffic that may not contain references to the personalized entities of users. For example, the personalized utterances could look like ``alexa make a phone call to john doe". These two data sources are combined to create the final training data set. Samples from the personalized data source are added until a predetermined ``personalized" fraction is reached. Users typically have a few thousand records of personalized "contact name" entities, each with a first and last name.



\begin{table}[]
    \centering
    \caption{Used training data sources, testing and validation datasets.}
    \label{tab:datasets}
    \begin{tabular}{c|l}
        Dataset Name & Hours of Audio \\
        \hline
        Personalized named entities training set & 131\\
        General training set & 91,656\\
        Personalized validation set & 40\\
        Personalized test set & 43\\
        General test set & 141
    \end{tabular}
\end{table}

\subsection{Models}

\subsubsection{First Pass Model}
We use a recurrent neural network transducer (RNN-T) based first pass model that also employs neural personalization~\cite{sathyendra2022contextual} and is supplemented with a shallow fusion n-gram model. The first pass score consists of two parts:
\begin{equation}
    u_i = \gamma \log P_{\mathrm{CA-RNN-T}} + \epsilon \log P_{\mathrm{SF}}, 
\end{equation} where $P_{\mathrm{CA-RNNT}}$ is the contextual-aware RNN-T model~\cite{sathyendra2022contextual} probability; $P_{\mathrm{SF}}$ is the shallow fusion n-gram probability; $\gamma$ and $\epsilon$ are interpolation weights. The context-aware RNN-T can improve the word error rate upto $\sim 35\%$ and the shallow-fusion by additional $\sim 5\%$ compared with a standard RNN-T model~\cite{sathyendra2022contextual}.

\subsubsection{Rescoring Models}

We use two different rescoring BERT models in this study: (i) tiny BERT; and (ii) big BERT. Both BERT models use a sentence-piece tokenizer with a multilingual vocabulary of $\sim$153k tokens (trained on the MC4 dataset~\cite{2019t5}). Roughly $\sim$30k of these tokens are from English language. The hyper-parameters for the model architecture are outlined in Table \ref{tab:models}.


\begin{table}[h]
    \centering
    \caption{Summary of BERT model hyper-parameters.}
    \label{tab:models}
    \begin{tabular}{c|c|c}
        Attribute & Tiny BERT & Big BERT \\
        \hline
        Layer Parameters & $\sim 5M$ & $\sim 170M$ \\
        Embedding Parameters & $\sim 49M$ & $\sim 157M$ \\
        Hidden Size & $320$ & $1024$ \\
        Number of Layers & $4$ & $16$ \\
        Number of Attention Heads &$16$& $16$ \\
        Intermediate Layer Dimension &$1200$ & $3072$\\
        Dropout & $0.1$ & $0.1$
    \end{tabular}
\end{table}

Both BERT models were first pre-trained on the MC4 dataset [30], and then domain-adapted using internal in-domain data. The BERT models were used to initialize the baseline RescoreBERT models, followed by MWER-based training on the general training data~\cite{xu2022rescorebert}. For the proposed personalization approaches in Section \ref{sect:approaches}, we initialize each model with the baseline RescoreBERT from the previous step. 


\subsection{Training and Evaluation Protocol}
All the models used the public hugging-face implementation of BERT~\cite{DBLP:journals/corr/abs-1810-04805} (model parameters and tokenizer are different). DeepSpeed~\cite{rajbhandari2020zero} was used for training and evaluating all models. Batch size was 256 and 128 for tiny and large BERT models, respectively. All experiments used a learning-rate decay policy and an Adam optimizer with default parameters. The initial learning rate was $10^{-5}$ for the tiny BERT experiments and $10^{-6}$ for the big BERT model experiments. An early stopping strategy, based on the personalized validation set, was applied (see Table \ref{tab:datasets}).



\section{Results and Discussion}

Relative word error rate reduction (WERR) numbers using vanilla RescoreBERT models, compared to a baseline without a rescoring pass, are shown in Table \ref{tab:baseline}. Using Oracle rescoring (e.g. picking the hypothesis with the smallest edit distance from truth) improves the first pass significantly, demonstrating the bound on WER improvements on both test sets.



RescoreBERT (see Fig. \ref{fig:illustration}a) improves on the general test set, but degrades the personalized test set. This is likely due to the fact that the first pass is already context aware, while the RescoreBERT models are not. The larger RescoreBERT amplifies this characteristic by causing a more pronounced degradation on the personalized test set. A fine-tuned RescoreBERT model reduces degradation but fails to match the first-pass WER on the personalized test set.


\begin{table}[h]
    \centering
    \caption{Baseline WERR results of rescoring models in the personalized and general test sets relative to the first pass (no rescoring). }
    \label{tab:baseline}
    \begin{tabular}{c|c|c}
        Model  & Personalized & General \\
        \hline
        Oracle& -57\%&-58\%\\
        Tiny RescoreBERT& +3.9\%&-5.3\%\\
        Big RescoreBERT& +4.8\%&-7.1\%\\
        Tiny RescoreBERT (fine-tuned)&+2.5\%&-5.3\%\\
        Big RescoreBERT (fine-tuned)&+1.7\%&-6.8\%
    \end{tabular}
\end{table}

\begin{table}[h]
    \centering
    \caption{Rescoring model WERR results in personalized and general test sets. Upper section: WERR relative to tiny RescoreBERT. Lower section: WERR relative to the big RescoreBERT. Negative sign corresponds to WER reduction and positive sign WER increase. Percentage sign in brackets in the model name denotes personalized training data fraction.}
    \label{tab:results}
    \begin{tabular}{c|c|c}
        Model  & Personalized & General \\
        \hline
        Untrained Prompt&-7.1\%&0\%\\
        Frozen Late&-4.5\% &0\%\\
        Frozen Early&-4.9\%&0\%\\
        Trained Late&-13\%&+3.9\%\\
        Trained Early&-13\%&+3.8\%\\
        Trained C-A&-14\%&+5.1\%\\
        \hline
        \hline
        Untrained Prompt& -12\%&+3.0\%\\
        Frozen Early&-6.1\%&0\%\\
        Trained Early (1\%)&-12\%&-0.7\%\\
        Fine-tuned Prompt (1\%)&-13\%&+0.4\%
    \end{tabular}
\end{table}


By applying natural language prompts without any training to RescoreBERT models (see Table \ref{tab:results}), WER improves significantly over the first pass with some degradation on the general test set. We hypothesize that this is particularly effective for the RescoreBERT model. The first-pass model penalizes longer utterances by assigning higher scores. Due to MWER training, the rescoring model learns to counter this by predicting a high-ranking score for longer utterances. By increasing the length of the hypothesis with adding semantically correct prompts, we believe that the hypothesis will receive better scores, thus getting ranked better than other hypotheses. We tested this anecdotally by trying out a variety of semantically correct and incorrect prompts and observed the same expected behavior.


Both early and late fusion gazetteers require additional training for the slot embeddings $E_s$ in Eqs. \eqref{eq:early}-\eqref{eq:late}. Freezing the weights of the vanilla RescoreBERT portion of the model, and training only the slot embeddings $E_s$ (``Frozen model"), improves the personalized-entity WER, with minimal degradation on the general test set compared with the baseline RescoreBERT model (see Table \ref{tab:results}). Early fusion yields a slightly larger improvement than the late fusion gazetteer. In spite of the training of the slot embeddings, the ``frozen model" gazetteers perform worse than prompting, which does not require training.


Training rescoring models with personalized data significantly improved WER on the personalized test set, but also caused degradation on the general test set. In addition, early fusion produced slightly better results than late fusion, so for the remainder of the experiments, we focus on early fusion. Cross-attention based encoder-decoder model gave the largest improvement on the personalized data, but also caused the largest degradation on the general data. This is partly explained by the larger number of trainable parameters.

\begin{figure}
    \centering
    \includegraphics[width=\linewidth]{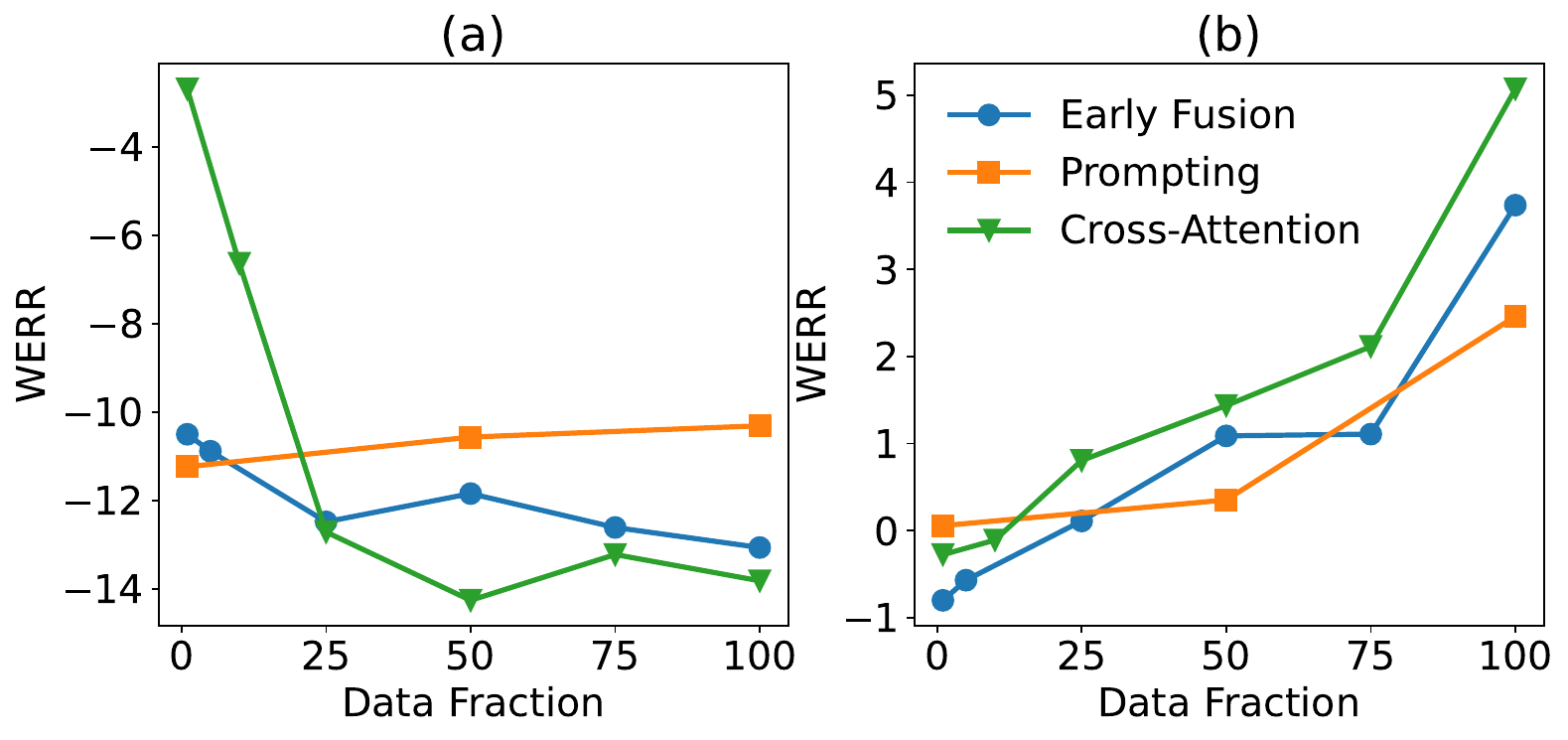}
    \caption{WERR respect to the tiny RescoreBERT model for early fusion gazetteer (blue line with circles), prompting (orange line with squares), and cross-attention (red line with triangles). Data fraction shown on the x axis is the fraction of personalized entity data in the used training data. Positive WERR denotes increase and minus sign reduction of WER respect to tiny RescoreBERT. (a) the personalized entity test set WERR for chosen approaches. (b) the general test set WERR for chosen approaches.}
    \label{fig:mixes}
\end{figure}

Fig. \ref{fig:mixes} shows training results for the models when general data is mixed into the training data. Using data mixing effectively mitigates the general test set degradation without significant loss of accuracy on the personalized test set with the exception of the cross-attention approach. The cross-attention model WER on the personalized test set became significantly worse when no general test set degradation was allowed. In the personalized test set, early fusion gazetteer performed slightly worse than cross-attention, but was less sensitive to data fraction. The early fusion model was also the only one that improved WER on the general test set over the baseline RescoreBERT, while also achieving over 10\% WERR on the personalized test set.


The bottom half of Table \ref{tab:results} presents the results of repeated experiments using the big BERT model. Prompting with the big RescoreBERT model proved very effective for personalization and provided a similar level of improvement as the trained models. However, it caused a larger degradation on the general test set than prompting with the tiny RescoreBERT. The trained early fusion gazetteer results improved in every aspect using a larger transformer.


\section{Conclusions}

We compared three different approaches to include personalized named entities in rescoring: (i) early and late fusion gazetteers; (ii) prompts; and (iii) cross-attention based encoder-decoder model. The effectiveness of these approaches was explored using internal data supplemented with personalized named entities. 
When trained with personalized data, all the approaches could deliver over 10\% WERR improvement over the baseline RescoreBERT model for the personalized test set. However, this leads to degradation in the general test set. Mixing some fraction of general data did address this issue for prompting and gazetteers with small loss of performance in the personalization test set. Cross-attention was more sensitive to the data fraction and we did not find a data mixing ratio that would yield good performance in both test sets. We also found that prompting gave 7\% improvement over the baseline WER without any training. This could be particularly useful for cases where training data is limited or not available. gazetteers proved to be the best all around approach as it was the only model that could give over 10\% improvement in the personalization while improving the general data test set at the same time. Both the gazetteers and prompting approaches performance become better with larger model size suggesting that they are scalable.

\bibliographystyle{IEEEtran}
\bibliography{mybib}

\end{document}